\documentclass[prl,twocolumn,floatfix,reprint,showpacs]{revtex4}
\usepackage{graphicx}
\usepackage{amsmath}
\usepackage{color}
\usepackage[usenames,dvipsnames]{xcolor}
\usepackage[colorlinks=true,citecolor=Cerulean,linkcolor=magenta,urlcolor=Cerulean]{hyperref}

\begin{document}

\title{Shortcuts to adiabaticity in the strongly-coupled regime: nonadiabatic control of a unitary Fermi gas}

\author{Shujin Deng$^1$, Pengpeng Diao$^1$, Qianli Yu$^1$,
Adolfo del Campo$^2$,
Haibin Wu$^{1*}$}

\affiliation{$^1$State Key Laboratory of Precision Spectroscopy, East China Normal University, Shanghai 200062, P. R. China}

\affiliation{$^2$Department of Physics, University of Massachusetts, Boston, MA 02125, USA}
\def\r{{\rm r}}
\def\G{\Gamma}
\def\L{\Lambda}
\def\la{\lambda}
\def\g{\gamma}
\def\al{\alpha}
\def\s{\sigma}
\def\e{\epsilon}
\def\k{\kappa}
\def\ve{\varepsilon}
\def\l{\left}
\def\r{\right}
\def\te{\mbox{e}}
\def\d{{\rm d}}
\def\t{{\rm t}}
\def\K{{\rm K}}
\def\N{{\rm N}}
\def\H{{\rm H}}
\def\la{\langle}
\def\ra{\rangle}
\def\om{\omega}
\def\Om{\Omega}
\def\vep{\varepsilon}
\def\wh{\widehat}
\def\tr{{\rm Tr}}
\def\da{\dagger}
\def\iz{\left}
\def\zi{\right}
\newcommand{\beq}{\begin{equation}}
\newcommand{\eeq}{\end{equation}}
\newcommand{\beqa}{\begin{eqnarray}}
\newcommand{\eeqa}{\end{eqnarray}}
\newcommand{\intf}{\int_{-\infty}^\infty}
\newcommand{\into}{\int_0^\infty}


\date{\today}

\begin{abstract}
Shortcut to adiabaticity (STA) guides the nonadiabatic dynamics of a quantum system towards an equilibrium state without the requirement of slow driving. We report the first demonstration of a STA in a strongly-coupled system, i.e., a 3D anisotropic Fermi gas at unitarity.  Exploiting  the emergent conformal symmetry, the time-dependence of the trap frequencies is engineered so that the final state in a nonadiabatic expansion or compression is stationary and free from residual excitations. The universal scaling dynamics is verified  both in the non-interacting limit and at unitarity.
\end{abstract}

\maketitle

The understanding of strongly-coupled quantum systems is an open problem at the frontiers of physics with widespread applications.
Prominent examples include the description of atomic nuclei, quark-gluon plasma, quantum liquids and other exotic phases of matter \cite{strongc}. In these systems, the presence of strong interactions challenges any attempt aimed at controlling their far-from equilibrium dynamics.

Shortcuts to adiabaticity (STA) have been recently been proposed as a general tool to tailor the nonadiabatic dynamics of quantum matter far from equilibrium \cite{STAreview}.
Their implementation has been particularly successful in the control of ultracold atomic clouds.
In this context, STA provide a mean to guide the  dynamics of an atomic cloud  in a time-dependent trap to connect different stationary equilibrium states without the requirement of slow driving.
The first experiment speeding up the expansion dynamics of an atomic thermal  cloud \cite{exp1} was soon followed by the demonstration of STA with Bose-Einstein condensates \cite{exp2} and low-dimensional quantum fluids with strong phase fluctuations \cite{exp3}. Parallel progress succeeded in engineering the nonadiabatic dynamics of few level systems \cite{expCD1,expCD2,expCD3} as well as continuous variable systems \cite{expCD4}.
However, all experimental reports to date focused on single-particle, non-interacting or weakly interacting systems.
In principle, theoretical results suggest that the control of general many-body systems via STA might be feasible \cite{DRZ12,Campbell15,OT16,DP16}. In practice, achieving this goal often requires knowledge of the spectral properties of the system \cite{Demirplak03,Berry09} that is hardly available under strong coupling. In such scenario, the existence of dynamical symmetries comes to rescue \cite{delcampo11,delcampo13,Deffner14,StringariSTA}.

In this Letter, we provide the first demonstration of shortcuts to adiabaticity in a strongly coupled system. To this end, we demonstrate the nonadiabatic many-body control of a three-dimensional unitary Fermi gas in a time-dependent anisotropic trap. Exploiting the emerging scale invariance at strong coupling, we implement superadiabatic expansion and compressions of the cloud and determine the reversibility of the dynamics via a quantum echo. Our  result establish the possibility of tailoring the dynamics of strongly-coupled quantum matter far form equilibrium.

\begin{figure}[t]
\begin{center}
\includegraphics[width=3.5 in]{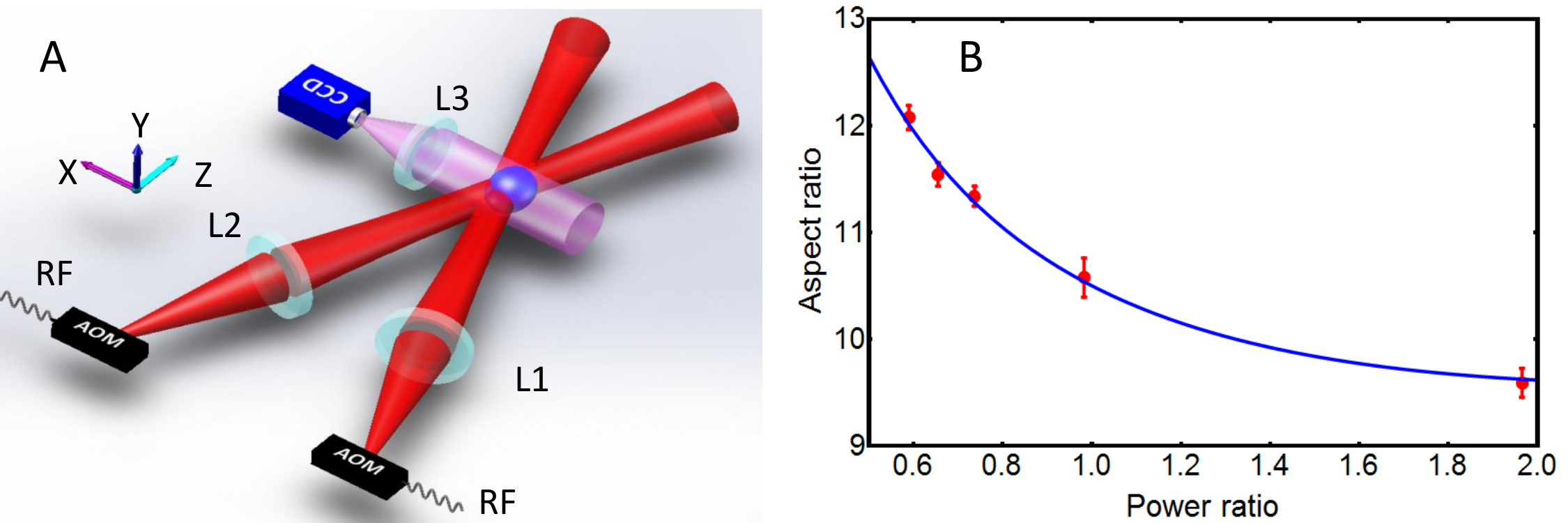}
\end{center}
\caption{ (A) Schematic  representation of experimental setup and  (B)  measured frequency aspect ratio  by controlling  power ratio of the two dipole-trap laser beams.  L1-L3, achromatic lenses; CCD, charge coupled device; AOM; acousto-optic modulator.
\label{fig:aspectratio}}
\end{figure}

\begin{figure*}
  \begin{minipage}{.31\textwidth}
    \includegraphics[angle=0,width=\textwidth,height=1.6 in]
    {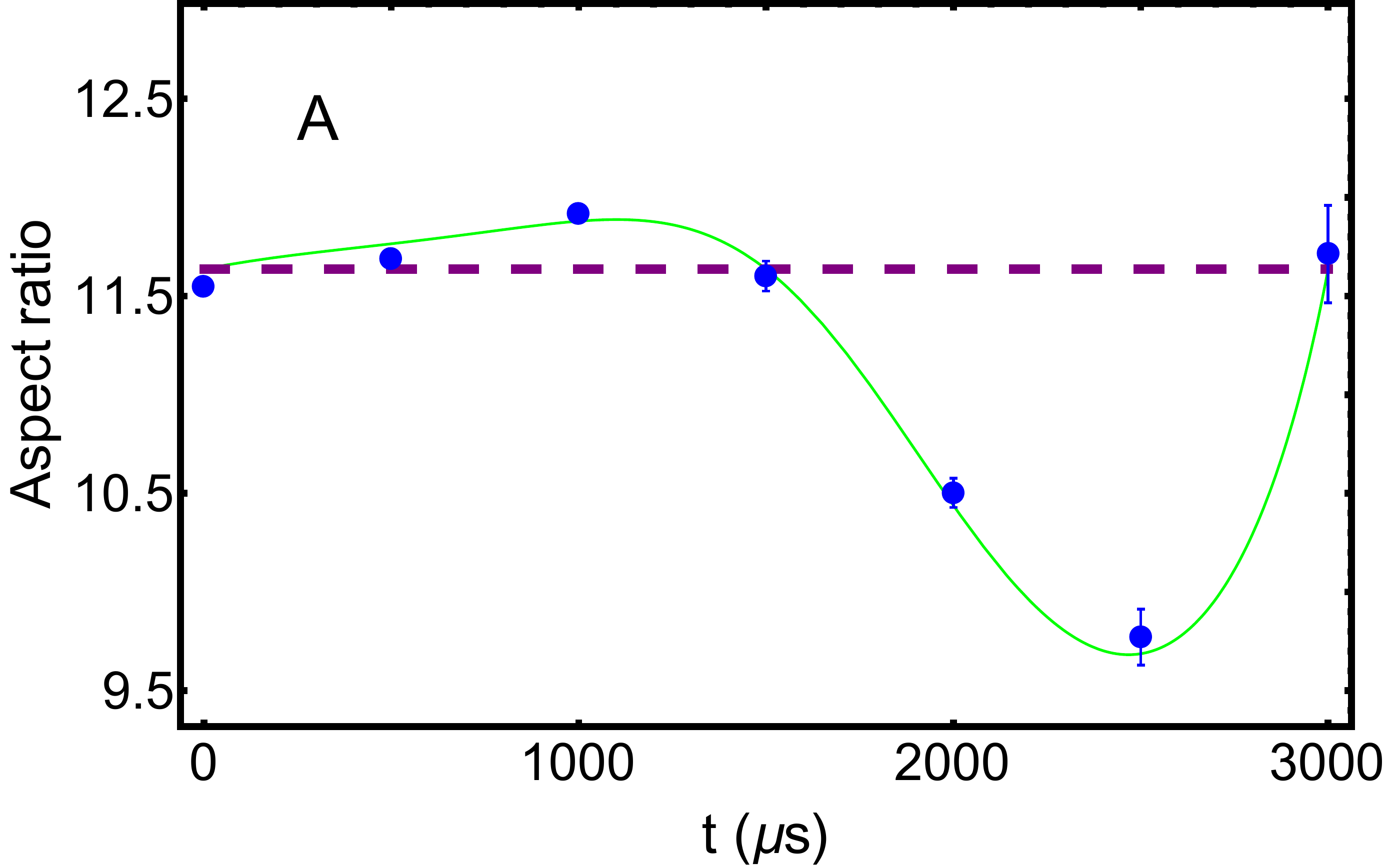}
  \end{minipage}
  \begin{minipage}{.31\textwidth}
    \includegraphics[angle=0,width=\textwidth,height=1.6 in]
    {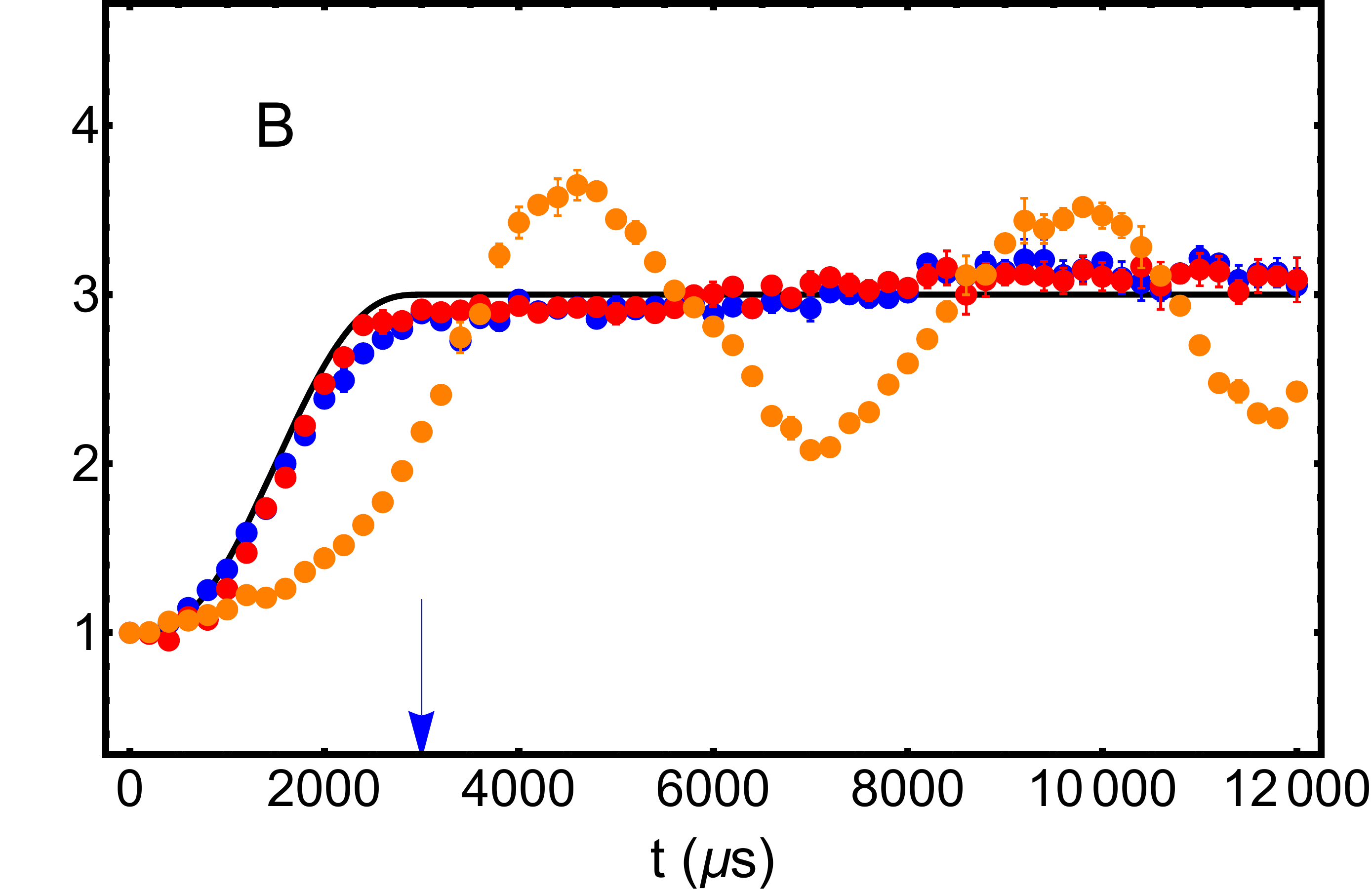}
  \end{minipage}
  \begin{minipage}{.31\textwidth}
    \includegraphics[angle=0,width=\textwidth,height=1.6 in]
    {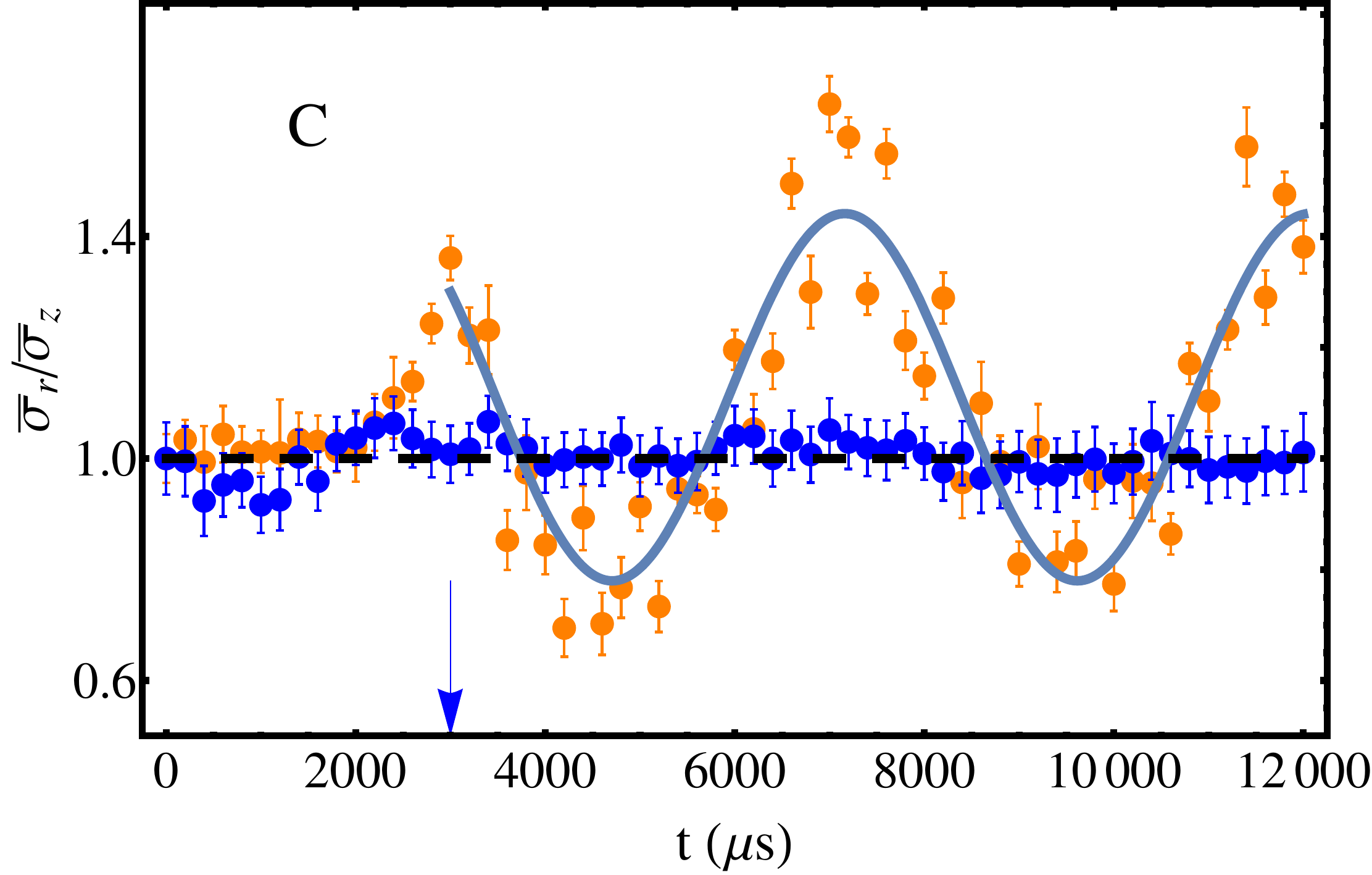}
  \end{minipage}
    \caption{\label{fig:STAdyn}
 {\bf Shortcut to the adiabatic expansion  of a Fermi gas at unitarity.}  The anisotropic 3D trap has an  initial aspect ratio $\omega_{0r}/\omega_{0z}=11.6$.  (A) The solid line (dots) is the theoretical (experimental) frequency aspect ratio for which an anisotropic STA is achieved  and the dashed line represents the initial frequency aspect ratio. (B) The blue dots and red dots are the measured mean square cloud size $\bar{\sigma}_z=\sigma_z/\sigma_{0z}$ and $\bar{\sigma}_r=\sigma_r/\sigma_{0r}$, respectively, for the STA trajectory. The black solid curve is the calculated mean square cloud size for STA. The orange dots are the axial mean square cloud size $\sigma_z/\sigma_{0z}$ for the non-STA trajectory where the trap frequency is linearly ramped. (C) The dynamics of $\bar{\sigma}_r/\bar{\sigma}_{z}$ for the STA (blue dots) and non-STA (orange dots) trajectory. The black dashed line denotes for $\bar{\sigma}_r/\bar{\sigma}_{z}=1$. The solid curve is a sinusoid fit for the data after $t_f$ to get the frequency of the collective mode.  The blue arrows denote for $t_f=3\,ms$. Here, $b_f=3$. Error bars represent the standard deviation of the statistic.
}
\end{figure*}
The unitary limit is reached in an ultracold Fermi gas at resonance. The divergent scattering length leads to the emergence of conformal invariance and  hydrodynamic behavior.  When the 3D unitary Fermi gas is harmonically trapped, the  many-body dynamics of the system becomes scale invariant and the evolution of the cloud is dictated by  the following coupled equations ~\cite{3DFermi}
\begin{equation}
\ddot{b}_j=\omega_{0j}^2/[b_j(b_xb_yb_z)^{2/3}]-\omega_j^2 b_j, \quad j=x, y, z
\label{b3DUnitaryEq}
\end{equation}
where $b_j(t)=\sigma_j(t)/\sigma_{0j}$  are time-dependent scaling factors satisfying the boundary conditions with $b_j(0)=1$ and $\dot{b}_j(0)=0$, and $\sigma_{0j}=\sigma_{j}(0)$ and $\sigma_{0j}=\sigma_{j}(0)$. As a result, the evolution of the cloud size  $\sigma_j(t)$ is dictated by the modulation of the trapping frequencies $\omega_j$.  By contrast, in the non-interacting Fermi gas that is collisionless, the equations decoupled
\begin{equation}
\ddot{b}_j=\omega_{0j}^2/b_j^3-\omega_j^2 b_j.
\label{b3DIdealEq}
\end{equation}
As a result,  the aspect ratio of the cloud evolves nonadiabatically and differs for  unitary and ideal Fermi gases under anisotropic harmonic confinement.  A well-known example is the anisotropic expansion associated with hydrodynamics behavior in a strongly-interacting degenerate Fermi gas~\cite{anisotropicexpansion}.

Here, following recent studies  \cite{Lobo08,delcampo11}, we implement the proposal by Papoular and Stringari \cite{StringariSTA} and experimentally demonstrate how to control the far-from-equilibrium  dynamics of a 3D unitary Fermi gas via a STA. By engineering the frequency aspect ratio of the trap as
 \begin{equation}
\omega_j^2(t)=\frac{\omega^2_{0j}}{b^4}-\frac{\ddot{b}}{b},
\label{frequencyEq}
\end{equation}
both unitary and ideal Fermi gases obey the same scaling dynamics, in which a single scaling factor $b$  suffices to completely describe the evolution of the cloud.  As a result, the dynamics of the system can be manipulated via  STA, speeding up the adiabatic transfer between two  many-body stationary states. The scale-invariant dynamics relies on the  conformal symmetry that we demonstrate in both an ideal and unitary Fermi gas.

Our experiment probes the dynamics of an anisotropically-trapped unitary quantum gas of a balanced mixture of $^6$Li fermions in the lowest two hyperfine states  $|\uparrow\rangle\equiv|F=1/2, M_F=-1/2\rangle$ and $|\downarrow\rangle\equiv|F=1/2, M_F= 1/2\rangle$. The experimental setup is similar to that  Ref.~\cite{Wu1,Wu2}. Fermionic atoms are loaded into a cross-dipole trap used for evaporative cooling. The resulting potential has a cylindrical symmetry around $z$ axis, as shown in Fig. \ref{fig:aspectratio}A. In an anisotropic trap, the typical frequency ratio $\omega_{0r}/\omega_{0z}$ is about 11.6.  A Feshbach resonance is used to tune the interaction of the atoms either to the non-interacting regime with the magnetic field $B=528\,$G or to the unitary limit with $B=832\,$G. The system is initially prepared in a stationary state with the trap depth fixed at $5\%\,U_0$ where $U_0$ is the full trap potential depth. The initial energy of Fermi gas at unitary is $E=0.8\,E_F$, corresponding to the temperature $T=0.25\,T_F$, where $E_F$ and $T_F$ are the Fermi energy and temperature of an ideal Fermi gas, respectively.  Note that the roundness of the optical trap beams and their alignment with the axes of the magnetic potential are crucial to this end \cite{SM}.  Any deviation from cylindrical symmetry and harmonic confinement due to misalignment, optical aberrations, or gravity could causes the excitation of the gas. The time dependence of $b(t)$ is engineered as a polynomial
\begin{equation}
b=1+(b_{f}-1)\bigg[10 (t/t_f)^3-15 (t/t_f)^4+6 (t/t_f)^5\bigg],
\label{bfactorEq}
\end{equation}
where $t_f$ is the transferring time to another stationary state and $b_{f}=b(t_f)$ is set by the ratio of  the initial and final target trap frequencies; $\omega_{fj}/\omega_{0j}=1/b^2(t_f)$ \cite{SM}.  The trap frequency is lowered by decreasing the laser intensity according to Eq.~(\ref{frequencyEq}), and trap anisotropy is precisely controlled by the power ratio of the two trap beams, see Figure \ref{fig:aspectratio}.
 Finally, after a time of evolution $t_f$ in the time-dependent trap, the trap beams are completely turned off and the cloud is probed via standard resonant absorption imaging techniques after a time-of-flight
 expansion time $t_{\rm tof}=200\,\mu s$. Each data point is an averaged over $5$ shots taken with identical parameters. The time-of-flight density profile along the axial (radial) direction is fitted by a Gaussian function, from which we obtain $\sigma_{z,obs}$($\sigma_{r,obs}$). $\sigma_{z,obs}$ ($\sigma_{r,obs}$) is related to the in-situ cloud size by a scale factor which can be obtained by either hydrodynamic or ballistic expansion equation with the time-of-flight time $t_{tof}$ \cite{SM}, from which  the mean
 cloud size $\sigma_z$ ($\sigma_r$) can be obtained.

The STA dynamics of a Fermi gas in the unitary regime and driven by the time-dependent anisotropic 3D trap is analyzed in Fig. \ref{fig:STAdyn}. Over the short time, the Fermi gas initially prepared in an equilibrium follows the engineered STA trajectory  and reaches a new stationary state where trapping frequencies are nine times as small as their initial values. The mean axial and radial cloud sizes share the same dynamics, as shown in Fig. \ref{fig:STAdyn}B. The evolution matches closely the theoretical prediction for the anisotropic STA trajectory and demonstrate  that the expansion of the unitary Fermi gas obeys the scaling law of Eq.~(\ref{frequencyEq}), which is consistent with the coupled Ermakov-like equations ~(\ref{b3DUnitaryEq}) when the scaling factors are tailored via STA to be isotropic $b(t)=b_j(t)$ for $j=x,y,z$. The residual heating rate of the cloud size is very small ($7.7\times10^{-4}\mu m/\mu s$) and negligible. The remarkable feature of the STA dynamics with new scaling solution is that $\sigma_{r}/\sigma_{z}$ remains constant when the gas is expanded by changing the trap aspect ratio, see blue dots in Fig. \ref{fig:STAdyn}C.  The STA evolution is contrasted with a typical nonadiabatic trajectory induced by a linear ramp of the trap frequency from the initial axial value $\omega_{iz}=2\pi\times 1324.7$ Hz to final one $\omega_{fz}=2\pi\times 147.2$ Hz  in $3\,$ms. The corresponding values of the mean axial cloud  $\sigma_z$ are shown in Fig. \ref{fig:STAdyn}B. Following the ramp, the Fermi gas is found far from equilibrium and the excitation of the breathing mode leads to an oscillatory behavior of the cloud size. The measured oscillation frequency is $2\pi\times 204.7$ Hz, which is determined by the collisions of atoms and very close to the frequency $\sqrt{12/5}\omega_{fz}$ of the axial collective mode  of Fermi gas at unitary Fermi gas for the final trap depth.

\begin{figure}[t]
\begin{center}
\includegraphics[width=0.8\linewidth]{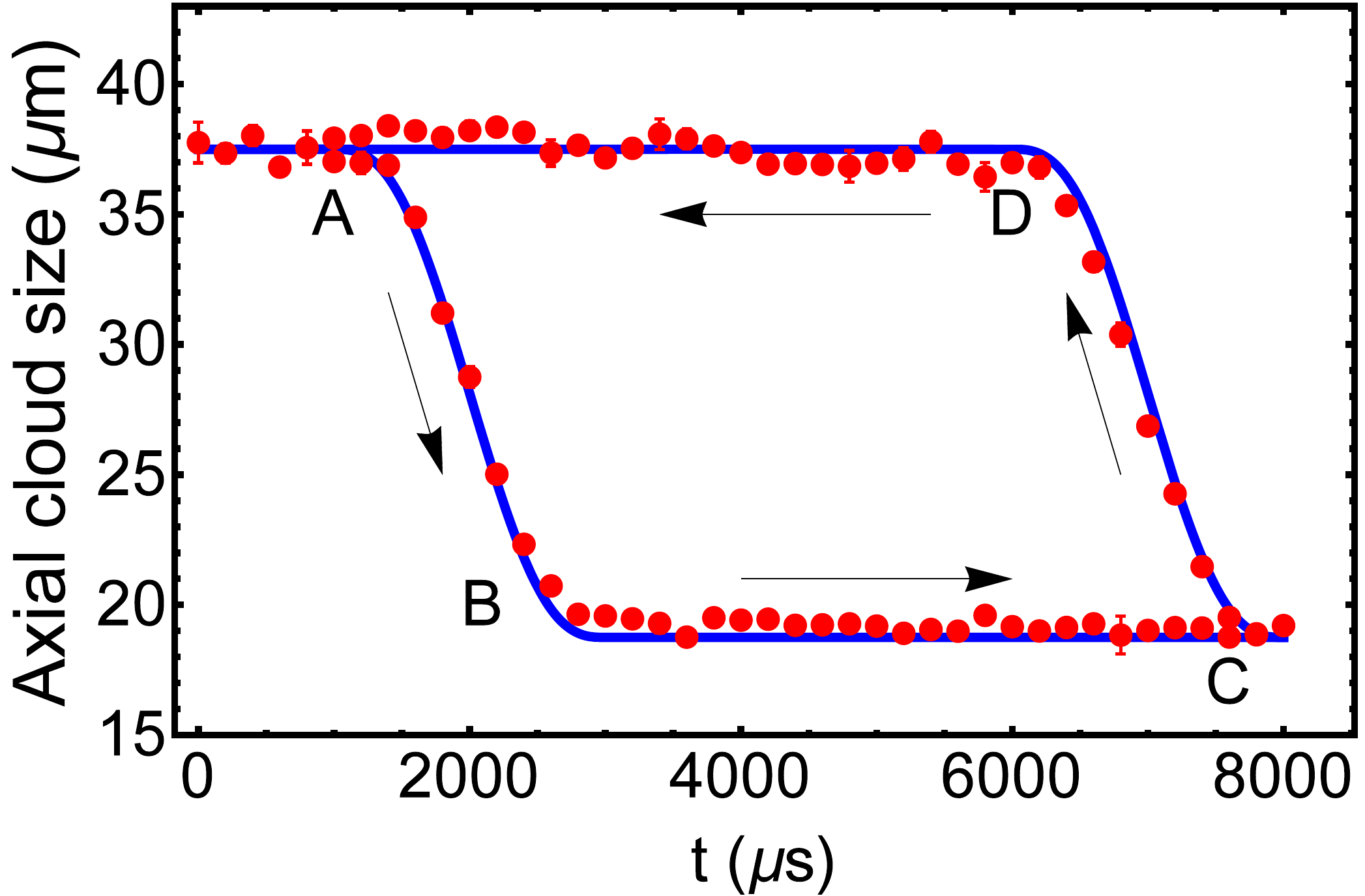}
\end{center}
\caption{{\bf Quantum echo adiabaticity test.} A sequence of  a STA expansion ($C\rightarrow D$), two isentropic processes($B\rightarrow C$ and $D\rightarrow A$, adiabatically holding the gas) and one STA compression process ($A\rightarrow B$). The solid curve is the theoretical prediction. Error bars represent the standard deviation of the statistic.
\label{fig:STAloop}}
\end{figure}

Armed with the ability to control nonadiabatic expansions and compressions via STA, we next text adiabaticity via a quantum echo \cite{QZ10} using the width of the cloud as a figure of merit.
The evolution of the mean axial size of the cloud  along the cycle is shown in Fig. \ref{fig:STAloop}.
 The initial system is prepared in a stationary state. The Fermi gas subsequently undergoes a  compression via a STA to reach a desired target stationary state in 3ms ($A\rightarrow B$) with no residual excitations. The gas is then adiabatically hold for 3ms ($B\rightarrow C$) before it evolves via a STA for an expansion ($C\rightarrow D$) which implements the inverse transformation of the previous STA compression. Finally there is an another isentropic process and the gas is hold for 3ms ($D\rightarrow A$).  Upon completion of the quantum echo cycle, we found that the dynamics is frictionless and the final mean axial cloud size matches closely the initial value.  Therefore it is proven that, if $\omega(t)$ is an STA expansion satisfying Eq.~(\ref{frequencyEq}) with the scaling parameter $b(t)$, then $\omega(t_f-t)$ is another STA compression. The STA process is time-reversible and the entropy indeed conserves in such transformations.

\begin{figure}[t]
\begin{center}
\includegraphics[width=0.7\linewidth]{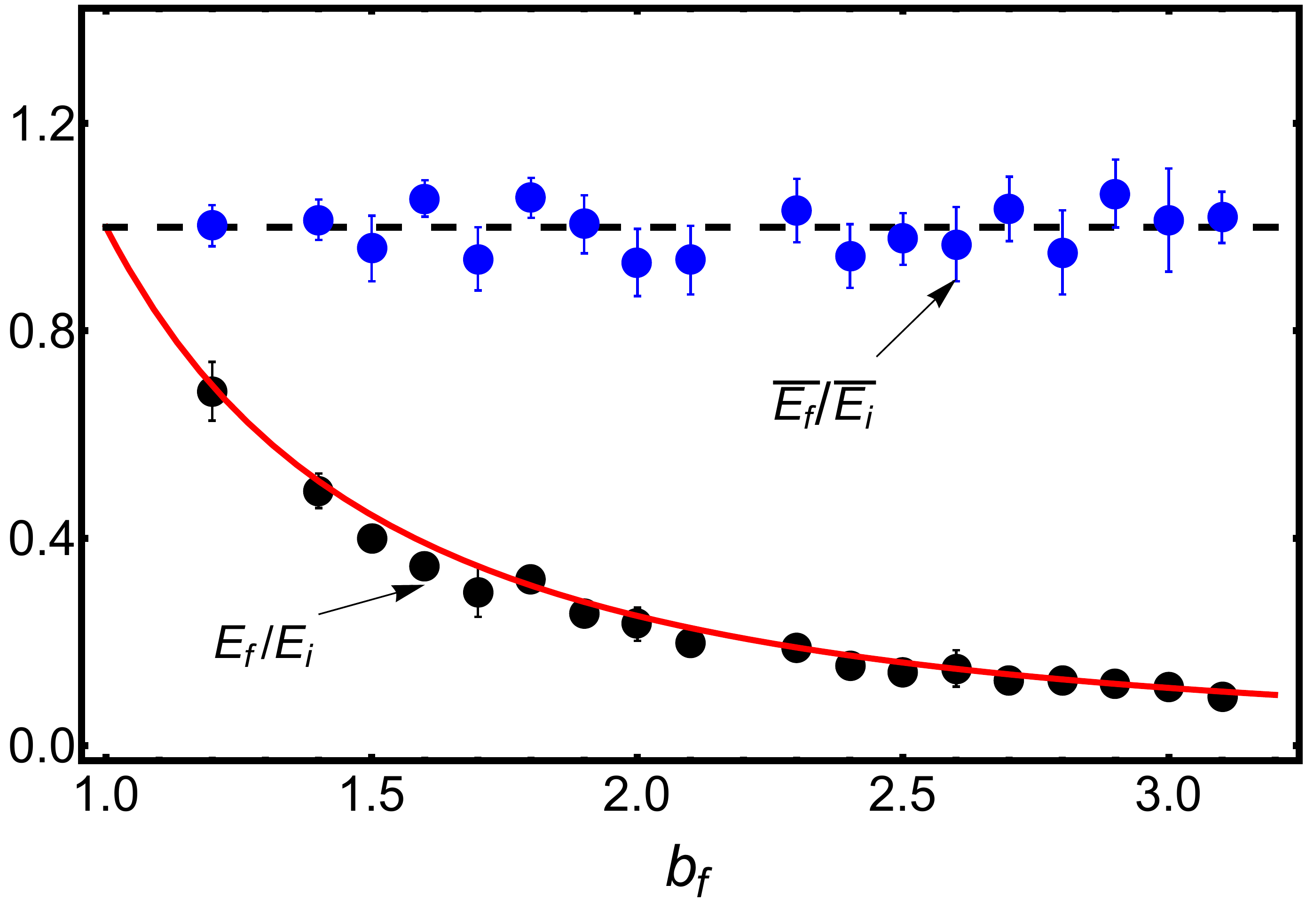}
\end{center}
\caption{{\bf Scaling law  of  nonadiabatic mean energy.} The black dots are the ratio of  the total energy of the target stationary state $E_f$ and the initial stationary state $E_i$ as a function of $b_f$. The blue dots are the dimensionless energy $\bar{E}_f/\bar{E_i}$  with different $b_f$, where $\bar{E}_f=E_f/E_{Ff}$ and $\bar{E}_i=E_i/E_{Fi}$. $E_{Fi}$($E_{Ff}$) is the Fermi energy of the initial(target) stationary state. The red solid curve is the theoretical prediction, $E_f/E_i=1/b_f^2$, and black dashed line represents $\bar{E}_f=\bar{E_i}$. Error bars represent the standard deviation of the statistic.
\label{fig:energy}}
\end{figure}

\begin{figure*}[t]
  \begin{minipage}{.45\textwidth}
    \includegraphics[angle=0,width=0.8\textwidth]
    {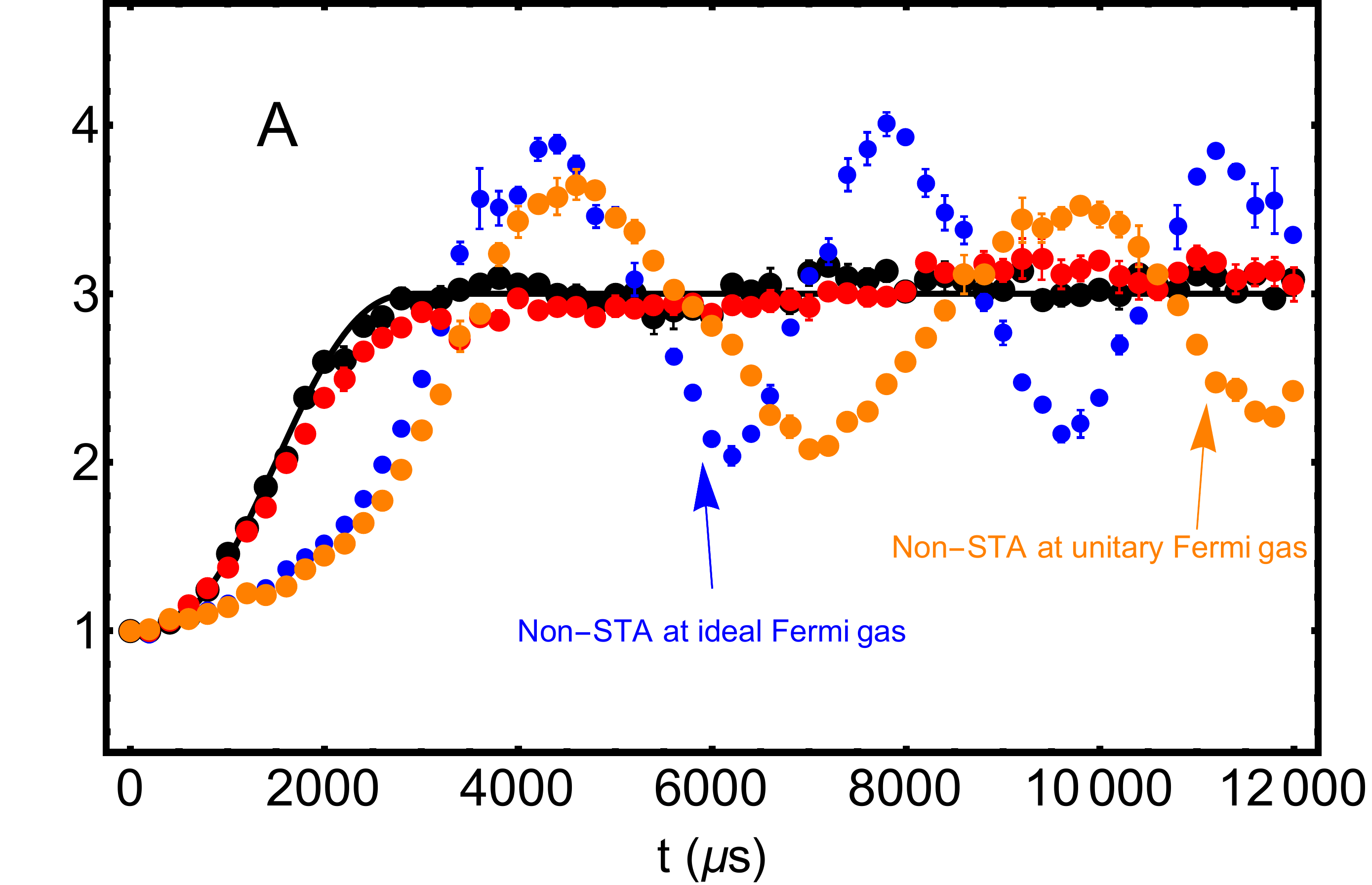}
  \end{minipage}
  \begin{minipage}{.45\textwidth}
    \includegraphics[angle=0,width=0.8\textwidth]
    {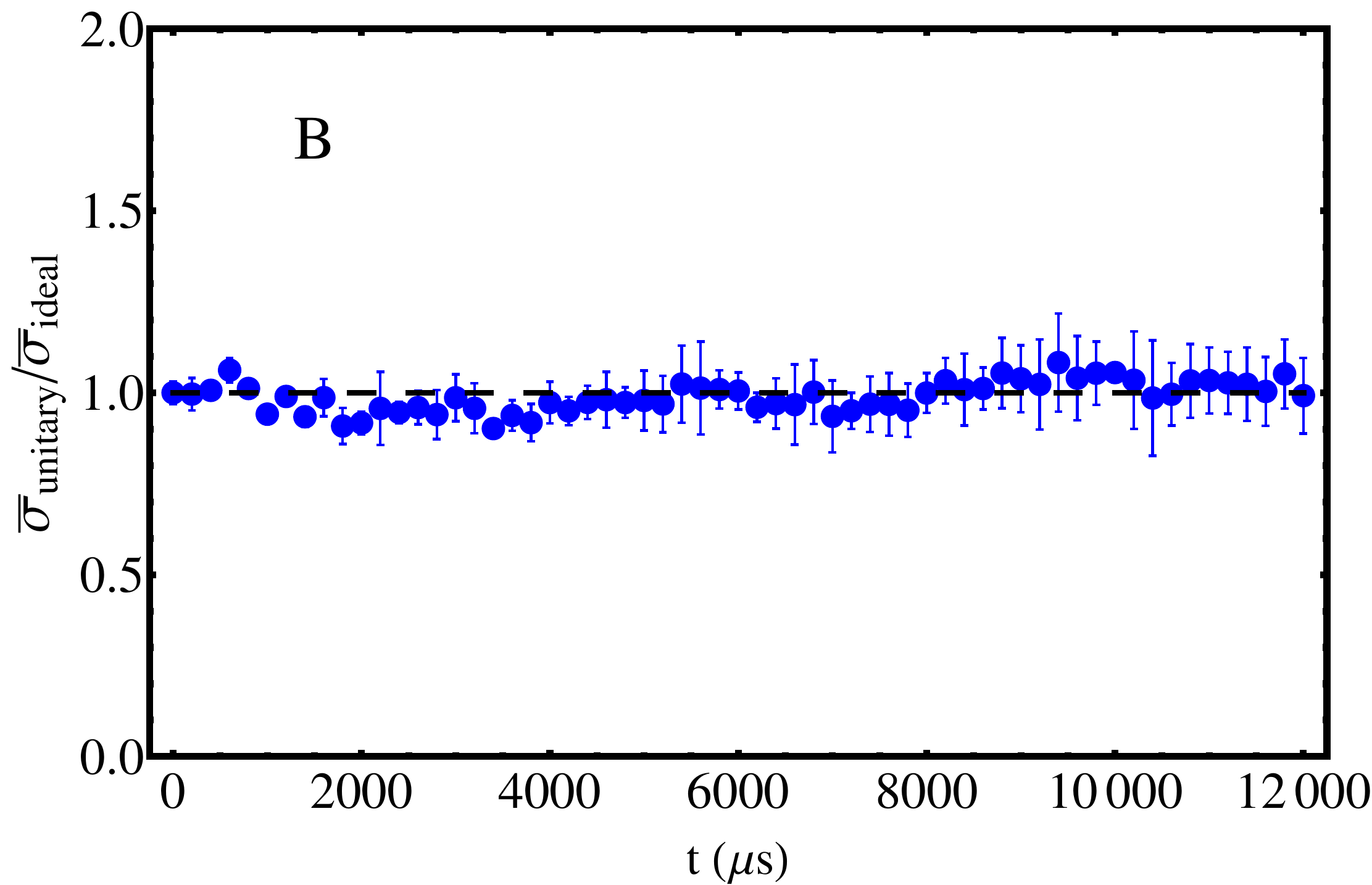}
  \end{minipage}
       \caption{\label{fig:universal}
    {\bf Compared STA dynamics for unitary and  ideal Fermi gases.} (A) The red dots and black dots are the STA trajectories  for the unitary Fermi gas at $B=832$ G and an ideal Fermi gas at  at $B=528$ G, respectively. Blue and orange dots are  for non-STA trajectories.  The black curve is the theoretical prediction based on Eq.~(\ref{frequencyEq}). (B) The dynamics of  $\bar{\sigma}_{unitary}/\bar{\sigma}_{ideal}$, where $\bar{\sigma}_{unitary,ideal}=\sigma(t)/\sigma_0$. The dashed black line is $\bar{\sigma}_{unitary}=\bar{\sigma}_{ideal}$. Error bars represent the standard deviation of the statistic.
}
\end{figure*}
To further characterize the nonadiabatic evolution along a STA, we next turn our attention to the mean-energy scaling dynamics.  The unitary  Fermi gases provide a non-relativistic fluid that is both scale and conformally invariant~\cite{scaleFermi}. The universal thermodynamics states that the relation of pressure $P$ and energy density $\epsilon$ is related by the same  equation of state $P = 2\epsilon/3$ as for an ideal gas. With a harmonic trap, the total energy per particle, $E/N = 3m\omega^2\langle\sigma\rangle^2$, is determined by the trap frequency $\omega$ and mean square cloud size $\langle\sigma\rangle^2$ with virial theorem. The ratio of the total energy between the target stationary state $E_f$ and the initial stationary state $E_i$ is investigated with different values of $b_f$ reached via STA.  Fig.~\ref{fig:energy} shows the excellent agreement between the theoretical prediction for  $E_f/E_i=1/b_f^2$ and the experimental data.  For large $b_f$ the energy of the final state could be greatly decreased. In the experiment,  the initial energy is decreased by an order of magnitude in 2 ms without residual excitations. Therefore, this anisotropic STA could be used as a new and efficient method to perform fast cooling for quantum gases.  We also notice that the dimensionless energy ratio $\bar{E_j}=E_j/E_{Fj}$ ($j=i,f$) does not change in such anisotropic  STA (see  Fig.~\ref{fig:energy}) although the external trap experiences a large change in the process, where $E_{Fi}$ and $E_{Ff}$ are the Fermi energy of a non-interacting two-state mixture of $^6$Li atoms in a harmonic trap of (geometric) mean frequency for the initial and target stationary state, respectively. In addition, in the experiment, the temperature and atoms number of unitary Fermi gas are controlled by adjusting the degree of evaporative cooling process. The results show that the STA dynamics is robust and insensitive to these parameters.

Finally, we demonstrate that these dynamics are universal. We first verify that both the unitary Fermi gas and  ideal Fermi gas obey this new scaling solution of Eq.~(\ref{frequencyEq}) for STA dynamics in the presence of anisotropic time dependent trap. The dynamics are shown in Fig.~\ref{fig:universal}. The red dots and blue dots are the STA trajectories with the mean axial cloud size for the unitary Fermi gas at $B=832$ G and  ideal Fermi gas at $B=528 G$, respectively, as shown in  Fig.~\ref{fig:universal}A. Both curves are collapsed into a single curve of designed dynamics of  Eq.~(\ref{frequencyEq}). For better show this point, the ratio $\bar{\sigma}_{unitary}/\bar{\sigma}_{ideal}$ of the dimensionless mean axial cloud size at unitary Fermi gas and ideal Fermi gas is plotted in Fig.~\ref{fig:universal}B, where $\bar{\sigma}_{unitary,ideal}=\sigma(t)/\sigma_0$. The data clearly demonstrate that $\bar{\sigma}_{unitary}/\bar{\sigma}_{ideal}\approx1$. The blue dots and orange dots are non-STA trajectories with the linear frequency ramp. For non-STA, the dynamical evolutions are different and their oscillation frequencies are related to breathing frequencies at the different regime with $\omega\approx 2\omega_{f}$ and $\omega\approx \sqrt{12/5}\omega_{f}$ for the ideal and unitary Fermi gas, respectively.

In conclusion, we have  provided the first experimental realization of a shortcut to adiabaticity in a strongly-coupled system. To this end, we have considered the scale-invariant dynamics of a 3D Fermi gas at unitarity in a time-dependent anisotropic harmonic trap. By designing the time-dependent frequency aspect ratio of the confinement, the nonadiabatic dynamics can be guided towards a stationary state in very short time scale.  Our work establishes the possibility to control quantum matter at strong coupling and far away from equilibrium.

We thank X. Chen and S. Stringari for helpful discussions. AdC thanks East China Normal University  for  hospitality during the completion of this work.
This research is supported by the National Natural Science Foundation of China (NSFC) (grant nos. 11374101 and 91536112), UMass Boston (project P20150000029279)  and the John Templeton Foundation.

\newpage
\widetext
\section{Supplemental Material}
\label{sec:supplement}

\subsection{Scale invariance of a unitary Fermi gas in an anisotropic harmonic trap and shortcuts to adiabaticity }

Scale invariance is a dynamical symmetry which has proved extremely useful in the study of ultra cold atomic gases in time-dependent harmonic traps.
In this context, an atomic cloud prepared at time $t=0$ is brought out of equilibrium by a modulation of the trap frequencies. Scale invariance  relates then correlation functions of the nonequlibrium state to those of the initial state, as was first discussed in \cite{CD96,KSS96} for Bose-Einstein condensates. However, scale invariance can be extended to both classical and quantum many-body  systems with $SU(1,1)$  as a dynamical symmetry group \cite{Gambardella75,Perelomov78}. The associated scaling laws are essential in the analysis of time of flight measurements \cite{GBD10}.  The possibility of engineering the scaling dynamics in many-body systems via shortcuts to adiabaticity was discussed in  to realize a quantum dynamical microscope \cite{delcampo11}.

For a superfluid Fermi gas, the conformal symmetry $SO(2,1)$ (isomorphic to $SU(1,1)$) is broken for any finite value of the scattering length and is only recovered as an emergent symmetry in the dilute collisionless limit describing a noninteracting Fermi gas and for a divergent scattering length, i.e., at unitarity.
Under isotropic harmonic confinement the scale-invariant dynamics of a unitary Fermi gas has been discussed in a number of theoretical works \cite{Castin04,Lobo08,StringariSTA}.  The possibility of using of STA in a unitary Fermi gas was first suggested by  Papoular and Stringari \cite{StringariSTA}.

A stationary quantum state $\Phi\left({\bf r}_1,\dots,{\bf r}_1;t=0\right)$ with chemical potential $\mu$ prepared at time $t=0$ in a trap with frequency $\om_0=\om(0)$ evolves following a modulation in time of $\om(t)$ according to the scaling law
\beqa
\label{sis}
\Phi\left({\bf r}_1,\cdots,{\bf r}_N;t\right)=
b^{-\frac{3N}{2}}\exp
\left[-i\frac{\mu\tau(t)}{\hbar}+i\frac{m\dot{b}}{2\hbar b}
\sum_{i=1}^N{\bf r}_i^2
\right]
\Phi\left(\frac{{\bf r}_1}{b},\cdots,\frac{{\bf r}_N}{b};0\right),
\eeqa
where
\beqa
\tau(t)=\int_{0}^t\frac{dt'}{b^2(t')}
\eeqa
 and the scaling factor is a solution of the Ermakov equation
\beqa
\label{EPE}
\ddot{b}+\om^2(t)b=\om_0^2/b^3
\eeqa
subjected to the  boundary conditions $b(0)=1$ and $\dot{b}(0)=0$. In the adiabatic limit $\dot{\om}(t)/\om(t)^2\ll1$ \cite{LR69}, the solution of the scaling factor is set by
\beqa
b_{\rm ad}(t)=\sqrt{\frac{\om_0}{\om(t)}}.
\eeqa
As a result, the adiabatic evolution under scale invariance simply reads
\beqa
\Phi_{\rm ad}\left({\bf r}_1,\cdots,{\bf r}_N;t\right)=
b_{\rm ad}^{-\frac{3N}{2}}\exp\left[-i\frac{\mu}{\hbar}\int_{0}^t\frac{dt'}{b_{\rm ad}^2(t')}\right]
\Phi\left(\frac{{\bf r}_1}{b_{\rm ad}},\cdots,\frac{{\bf r}_N}{b_{\rm ad}};0\right).
\eeqa
The adiabaticity condition can be removed via a STA in an arbitrarily short time $t_f$ by driving the trapping frequency from $\om_0$ to $\om(t_f)$.
By  imposing the boundary condition on the scaling factor
\beqa
b(t_f)=b_{\rm ad}(t_f), \quad \dot{b}(t_f)=0,
\eeqa
the time-evolving state (\ref{sis}) reduces to the corresponding stationary state at $\om(t_f)$, as the phase factor proportional to $\dot{b}/b$ vanishes. The required trapping frequency can be determined from the Ermakov solution by choosing any function $b(t)$ that satisfies the boundary conditions at $t=0,t_f$. To ensure a smooth modulation of the trapping frequency is generally convenient to add the supplementary boundary conditions  $\ddot{b}(0)=\ddot{b}(t_f)=0$. The polynomial ansatz discussed in the main body of the manuscript is one possible choice.

When the three-dimensional harmonic trap is anisotropic, the scaling dynamics is generally more complex \cite{Menotti02} as the scaling factors along different degrees of freedom take different values \cite{CD96}.
However, it still possible to use the scaling law  (\ref{sis}) whenever $b_j(t)=b(t)$ \cite{Lobo08}. In the adiabatic case this is automatically satisfied provided that the aspect ratio of the trap is keep constant during the modulation of the trapping frequency. In the nonadiabatic regime,  the Ermakov equation for each trapping frequency becomes
\beqa
\ddot{b}+\om_j^2(t)b=\om_{j0}^2/b^3, \quad j=x,y,z.
\eeqa
Note that these equations are decoupled an can be easily inverted to determine $\om_j(t)$ using a common choice of the evolution of the scaling factor $b(t)$  and taking into account the corresponding initial frequency $\om_{j0}=\om_j(0)$ along each axis.

\subsection{Experimental methods}

\begin{figure}[t]
\begin{center}
\includegraphics[width=150mm]{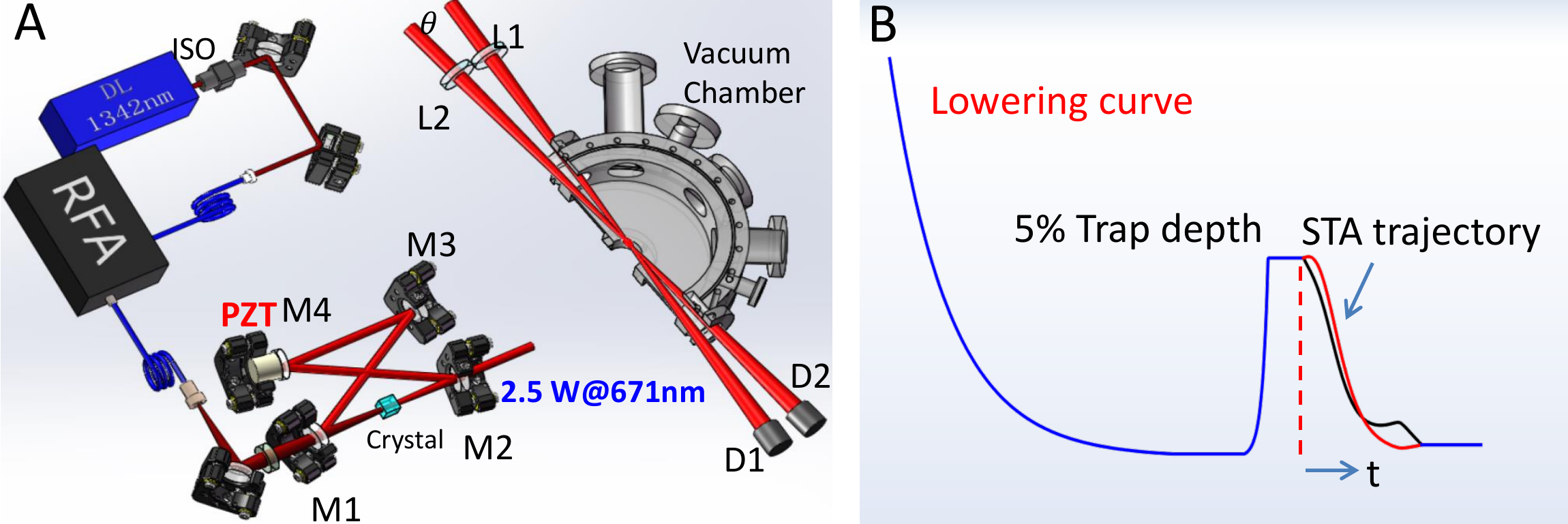}
\end{center}
\caption{The schematic of experimental setup (A) and the trap lowering curve (B).  RFA, Raman fiber amplifier; ISO, optical isolator; M1-M4, doubling-cavity mirrors; PZT, piezoelectric transducer; L1-L2, achromatic lenses; D1-D2, dumpers.
\label{fig:SFig1}}
\end{figure}

Our experimental setup was presented in \cite{Wu1,Wu2}. A large atom number magneto-optical trap (MOT) is realized by a laser system of 2.5-watts laser output with Raman fiber amplifier and intracavity-frequency-doubler, as shown in Fig.~\ref{fig:SFig1}(A).  After the loading stage, $5.2\times 10^9$ cold atoms is obtained.  Then MOT is compressed to obtain atoms of near Doppler-limited temperature of 280 $\mu K$ and a density of $10^{12}$ cm$^{-3}$. Following this compressed stage, the MOT gradient magnets are extinguished and repumping beams are switched off faster than the cooling beams. By optical pumping, a balanced mixture of atoms in the two lowest hyperfine states $|1\rangle\equiv|F=1/2, M=-1/2\rangle$ and $|2\rangle\equiv|F=1/2, M= 1/2\rangle$ is prepared.

The optical crossed dipole trap is consisted by a 200$\,$ W Ytterbium fiber laser at 1070$\,$nm. The resulting potential has a cylindrical symmetry along $z$ axis. The trap frequencies can be determined by
\begin{eqnarray}
\omega_x&=&\sqrt{\frac{4kU_1}{m\omega_1^2}+\frac{4kU_2}{m\omega_2^2}},\\
\omega_y&=&\sqrt{\frac{kU_1}{m}\bigg(\frac{1}{z_{R1}^2}+\frac{2}{\omega_1^2}\bigg)+\frac{kU_2}{m}\bigg(\frac{1}{z_{R2}^2}+\frac{2}{\omega_2^2}\bigg)+(X1^2+4X2)^{1/2}},\\ \omega_z&=&\sqrt{\frac{kU_1}{m}\bigg(\frac{1}{z_{R1}^2}+\frac{2}{\omega_1^2}\bigg)+\frac{kU_2}{m}\bigg(\frac{1}{z_{R2}^2}+\frac{2}{\omega_2^2}\bigg)-(X1^2+4X2)^{1/2}},
\end{eqnarray}
where $X1=\frac{kU_1}{m}\bigg(\frac{1}{z_{R1}^2}-\frac{2}{\omega_1^2}\bigg)-\frac{kU_2}{m}\bigg(\frac{1}{z_{R2}^2}-\frac{2}{\omega_2^2}\bigg)$ and $X2=4\frac{kU_1}{m}\bigg(\frac{1}{z_{R1}^2}-\frac{2}{\omega_1^2}\bigg) \frac{kU_2}{m}\bigg(\frac{1}{z_{R2}^2}-\frac{2}{\omega_2^2}\bigg)\cos^2\theta$. $z_{R1}$($\omega_1$) and $z_{R2}$ ($\omega_2$) are the Rayleigh lengths (waists) of two crossed dipole trap beams, respectively.  The crossed angle is about $\theta=19^0$.  $U_1$ and $U_2$ are the trap potential for each dipole beams. The trap frequencies can be controlled by the power and angle of two dipole laser beams using AOMs.

\begin{figure}[t]
\begin{center}
\includegraphics[width=120mm]{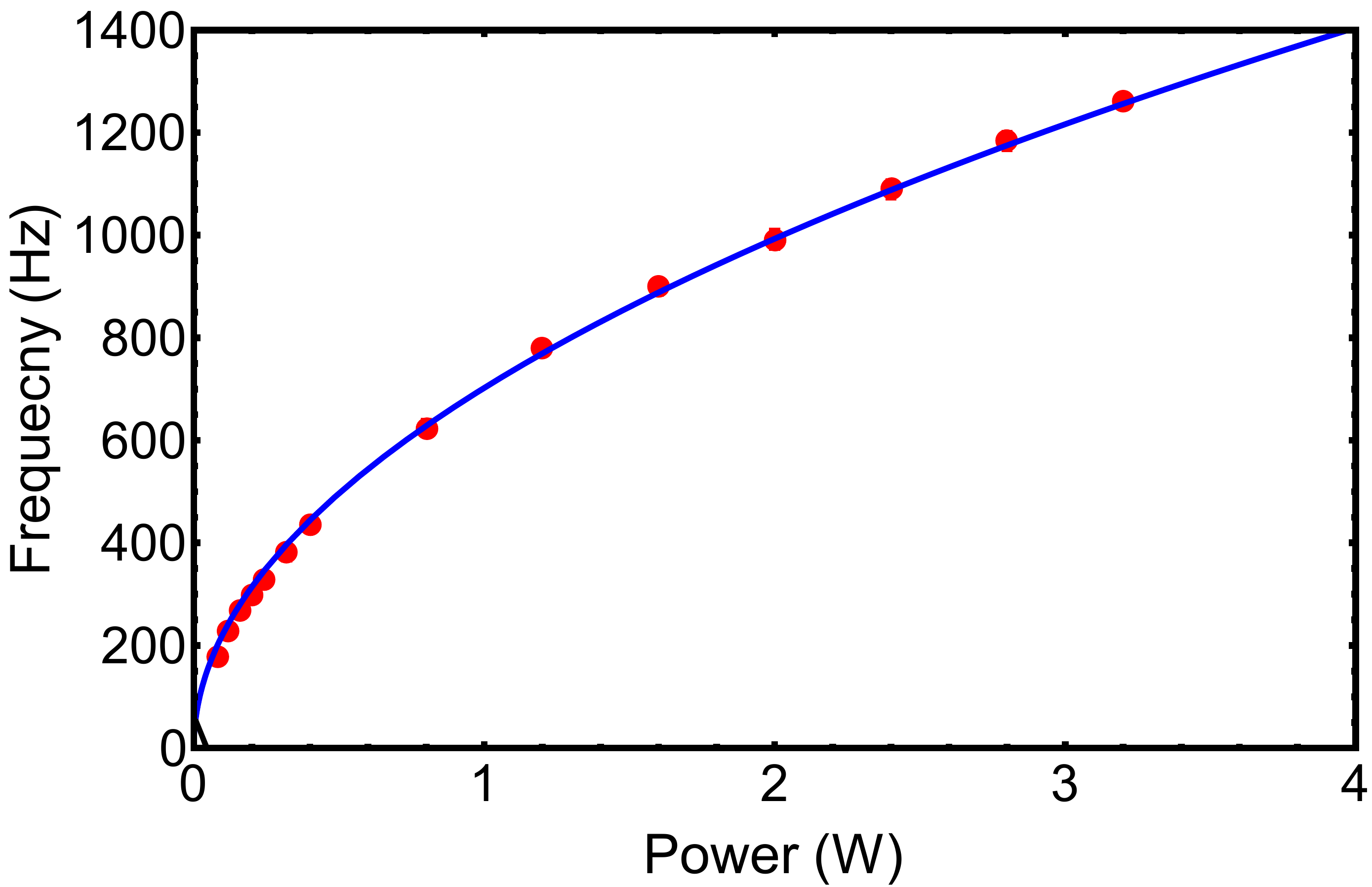}
\end{center}
\caption{The dependence of the frequency on the power of trap laser beams. The red dots are measured data  by parametric resonance. The blue solid curve is the calculation with the trap frequency to be proportional to $\sqrt{P}$.
\label{fig:SFig2}}
\end{figure}
The evaporative cooling is performed at Feshbach resonance of the magnetic field $B=832$ Gauss.  The trap is turned on 100 ms before the MOT compressed stage. After the atoms are loaded into the dipole trap we hold the atoms 200 ms on the trap and then forced evaporative cooling is followed by lowering the trap laser power, as shown in Fig.~\ref{fig:SFig1}B. First a simple exponential ramp [$U_0\exp(-t/\tau)$] is used as a lowering curve, where $U_0$ is the full trap depth. The time constant $\tau$ s is selected to control the trapping potential down to a variable value, which allowing us to vary the final temperature and atom number of the cloud. After the forced evaporative cooling  the dipole trap is recompressed to a value $5\% \,U_0$. Then the trap depth is held 0.5 s for the equilibrium. Then the frequency aspect ratio is controlled by adjusting the power ratio of two dipole laser beams and a STA trajectory is performed.   After the time $t$, the trap is turned off and a standard absorption image is following to extract the cloud profiles. In the analysis of the images we use a Gaussian function to fit the column density and get the mean cloud size.

\subsection{Determination of the scaling factor in time-of-flight measurements}
\label{sec:scaling}

The STA dynamical evolution is investigated by measuring the cloud radius after it is suddenly released from the trap.  The observed cloud size is then related to the initial cloud size inside the trap via the scaling  factor $b_{\rm tof}$.
To determine it for a unitary Fermi gas, we exploit  the hydrodynamic approach described in Ref. ~\cite{Johnscale}. We first note that due to the emergent conformal symmetry the density is given by
\begin{equation}
n({\mathbf{r}},t)=\frac{n_0(x/b_x,y/b_y,z/b_z)}{\Gamma},
\label{eq:density}
\end{equation}
where $b_i(t)$, $i=x,y,z$ are a time dependent scale factors. Here, $n_0$ denotes the density profile of the trapped cloud before the beginning of the Efimovian expansion. The function $\Gamma(t)\equiv b_xb_yb_z$ determines the scaling of the total cloud volume, that is independent of the spatial coordinates in the scaling approximation. With Eq.~(\ref{eq:density}), a velocity field is linearly dependent on the spatial coordinates, $v_i=x_i\,\dot{b}_i/b_i$.
Here $\langle x_i^2\rangle=\langle x_i^2\rangle_0\,b_i^2(t)$, and $\langle v_i^2\rangle=\langle x_i^2\rangle\,\dot{b}_i^2/b_i^2=\langle x_i^2\rangle_0\,\dot{b}_i^2(t)$, where $\langle x_i^2\rangle_0$ is the mean-square cloud radius of the trapped cloud along the $i$-direction just before releasing. Then, with these scaling assumptions and for the unitary Fermi gas, the fluid dynamics equations can be written as
\begin{equation}
\langle x_i^2\rangle_0\,b_i\,\ddot{b}_i=\frac{1}{Nm}\int d^3{\mathbf{r}}\,p-\frac{1}{m}\langle x_i \partial_i U_{\rm total}\rangle,
\label{eq:3.1}
\end{equation}
where the heating from the shear viscosity is neglected \cite{Caoscience}.

Before the Efimovian expansion, the balance of the forces arising from the pressure and the trapping potential yields
\begin{equation}
\frac{1}{N}\int d^3{\mathbf{r}}\,p=\frac{\langle{\mathbf{r}}\cdot\nabla U_{\text{total}}\rangle_0}{3\,\Gamma^{2/3}}.
\label{eq:3.2}
\end{equation}
Substituting Eq.~(\ref{eq:3.2}) into Eq.~(\ref{eq:3.1}), we get the evolution equations of the expansion factors,
\begin{equation}
\ddot{b}_i=\frac{\omega_{i0}^2}{\Gamma^{2/3}b_i}-\frac{\langle x_i \partial_i U_{\text{total}}\rangle}{m\langle x_i^2\rangle_0 b_i},
\label{eq:3.3}
\end{equation}
where
\begin{equation}
\omega_{i0}^2\equiv\frac{\langle x_i\partial_i U_{\text{total}}\rangle_0}{m\langle x_i^2\rangle_0}
=\frac{\langle{\mathbf{r}}\cdot\nabla U_{\text{total}}\rangle_0}{3m\langle x_i^2\rangle_0}.
\label{eq:3.4}
\end{equation}
The complete trapping potential is denoted by $U_{total}=U_{opt}+U_{mag}$ and includes  the optical trapping potential $U_{\rm opt}$ as well as  the residual magnetic potential $U_{\rm mag}$ from the Feshbach coils.
The later is $\sim$20Hz and can be ignored for the design of a STA.
For a harmonic trap potential, Eq.~(\ref{eq:3.4}) becomes
\begin{equation}
\ddot{b}_i=\frac{\omega_{i0}^2}{\Gamma^{2/3}b_i}-\omega_{\text{\rm mag}}^2 b_i-\omega_{i,\text{\rm opt}}^2(t) b_i,
\label{eq:3.5}
\end{equation}
where $\omega_{i0}^2$ is measured with the method of parametric resonance from the cloud profile and $\omega_{i,{opt}}$ is the optical trap frequency used to implement STA in the experiment. Therefore $b_i(t_{\rm tof})$ for flight time $t_{\rm tof}=200\mu s$ is determined by Eq.~(\ref{eq:3.5})  with the initial conditions $b_i(0)=1$ and $\dot{b}_i(0)=0$.

\subsection{The anisotropic harmonic trap}
In the experiment, the anharmonicity of the trap potential is undesirable to  realize the anisotropic STA.  Any deviation from cylindrical symmetry and harmonics trap potential owing to misalignment, optical aberrations could cause the obvious excitation of Fermi gas. For the harmonic trap potential, the trap frequency should be proportional to the square root of the power of the trapped laser beams. Parametric resonance
is used to measure the oscillation frequencies of weakly interacting atoms. The dependence of the frequency on the power is shown in Fig.~\ref{fig:SFig2}. For the high trap depth (high laser power), the trapped is essentially harmonic, i.e. anharmonic corrections are negligible. However, at the lower trap depth, the small laser power, the trap shows a relatively large anharmonicity. This is due to that the correct factor  for anharmonicity in the trapping potential is from $E_F/U_0$, where $E_f$ and $U_0$ are the Fermi energy and trap depth, respectively. In the realization of STA, we measure the frequencies and verify that the anharmonicity is smaller than 2$\%$.


\begin{thebibliography}{10}

\bibitem{strongc} A. Adams, L. D. Carr, T. Sch\"afer, P. Steinberg, J. E. Thomas,
\href{10.1088/1367-2630/14/11/115009}{New J. Phys. {\bf 14}, 115009 (2012).}

\bibitem{STAreview}
E. Torrontegui, S. Iban\~ez, S. Mart\'inez-Garaot, M. Modugno, A. del Campo, D. Gu\'ey-Odelin, A. Ruschhaupt, X. Chen, and J.G. Muga,
\href{10.1016/B978-0-12-408090-4.00002-5}{Adv. At. Mol. Opt. Phys. {\bf 62}, 117 (2013).}

\bibitem{exp1}
J. F. Schaff, X. L. Song, P. Vignolo, and G. Labeyrie, \href{https://doi.org/10.1103/PhysRevA.82.033430}{Phys. Rev. A {\bf 82}, 033430 (2010).}

\bibitem{exp2}
J. F. Schaff, X. L. Song, P. Capuzzi, P. Vignolo, and G. Labeyrie, \href{https://doi.org/10.1209/0295-5075/93/23001}{ EPL {\bf 93}, 23001 (2011).}

\bibitem{exp3}
W. Rohringer, D. Fischer, F. Steiner, I. E. Mazets, J. Schmiedmayer, M. Trupke, \href{https://doi.org/10.1038/srep09820}{Sci. Rep. {\bf 5}, 9820 (2015).}


\bibitem{expCD1} M. G. Bason, M. Viteau, N. Malossi, P. Huillery, E. Arimondo, D. Ciampini, R. Fazio, V. Giovannetti, R.  Mannella, O. Morsch,  \href{http://dx.doi.org/10.1038/nphys2170}{Nature Phys. {\bf 8}, 147 (2012).}

\bibitem{expCD2} J. Zhang, J. Hyun Shim, I. Niemeyer, T. Taniguchi, T. Teraji, H. Abe, S. Onoda, T. Yamamoto, T. Ohshima, J. Isoya, D. Suter,
\href{http://dx.doi.org/10.1103/PhysRevLett.110.240501}{Phys. Rev. Lett. {\bf 110}, 240501 (2013).}

\bibitem{expCD3} Yan-Xiong Du, Zhen-Tao Liang, Yi-Chao Li, Xian-Xian Yue, Qing-Xian Lv, Wei Huang, Xi Chen, Hui Yan, Shi-Liang Zhum, \href{http://dx.doi.org/10.1038/ncomms12479}{Nat. Commun. {\bf 7}, 12479 (2016). }

\bibitem{expCD4} S. An, D. Lv, A. del Campo, K. Kim,  \href{http://dx.doi.org/10.1038/ncomms12999}{Nature Commun. {\bf 7}, 12999 (2016).}



\bibitem{DRZ12} A. del Campo, M. M. Rams, and W. H. Zurek,
\href{http://dx.doi.org/10.1103/PhysRevLett.109.115703}{Phys. Rev. Lett. {\bf 109}, 115703 (2012).} 

\bibitem{Campbell15}
S. Campbell, G. De Chiara, M. Paternostro, G. Palma, and R. Fazio, \href{https://doi.org/10.1103/PhysRevLett.114.177206}{Phys. Rev. Lett. {\bf 114},177206 (2015).}
\bibitem{OT16}
M.  Okuyama and K. Takahashi, \href{https://doi.org/10.1103/PhysRevLett.117.070401}{Phys. Rev. Lett. {\bf 117}, 070401 (2016).}
\bibitem{DP16} D. Sels and A. Polkovnikov, \href{http://lanl.arxiv.org/abs/1607.05687}{arXiv:1607.05687 (2016)}.


\bibitem{Demirplak03} M. Demirplak, and Stuart A. Rice,
\href{http://dx.doi.org/10.1021/jp030708a}{J. Phys. Chem. A {\bf 107}, 9937 (2013).} 

\bibitem{Berry09} M. V. Berry,
 \href{http://dx.doi.org/10.1088/1751-8113/42/36/365303}{J. Phys. A: Math. Theor. {\bf 42}, 365303 (2009).} 


\bibitem{delcampo11} A. del Campo
\href{http://dx.doi.org/10.1103/PhysRevA.84.031606}{Phys. Rev. A {\bf 84},  031606(R) (2011).}
\bibitem{delcampo13}
A. del Campo,  \href{http://dx.doi.org/10.1103/PhysRevLett.111.100502}{Phys. Rev. Lett. {\bf 111}, 100502 (2013).}
\bibitem{Deffner14} S. Deffner, C. Jarzynski, and A. del Campo,
\href{http://dx.doi.org/10.1103/PhysRevX.4.021013}{Phys. Rev. X {\bf 4}, 021013 (2014).}
\bibitem{StringariSTA}
D. Papoular and S. Stringari, \href{https://doi.org/10.1103/PhysRevLett.115.025302}{Phys. Rev. Lett. {\bf 115}, 025302 (2015).}

\bibitem{3DFermi}
C. Menotti, P. Pedri, and S. Stringari, \href{https://doi.org/10.1103/PhysRevLett.89.250402}{Phys. Rev. Lett. {\bf 89}, 250402 (2002).}

\bibitem{anisotropicexpansion}
K. M. O'Hara, S. L. Hemmer, M. E. Gehm, S. R. Granade, J. E. Thomas, \href{https://doi.org/10.1126/science.1079107}{Science {\bf 298}, 2179 (2002).}

\bibitem{Lobo08}
C. Lobo and S. D. Gensemer, \href{https://doi.org/10.1103/PhysRevA.78.023618}{Phys. Rev. A {\bf 78}, 023618 (2008).}




\bibitem{Wu1}
S. Deng, P. Diao, Q. Yu, and H. Wu,  \href{https://doi.org/10.1088/0256-307X/32/5/053401}{Chin. Phys. Lett. {\bf 32}, 053401 (2015).}

\bibitem{Wu2}
S. Deng, Z. Shi, P. Diao, Q. Yu, H. Zhai, R. Qi, and H. Wu, \href{https://doi.org/10.1126/science.aaf0666 }{Science {\bf 353}, 371 (2016).}
\bibitem{SM} See Supplementary Material.

\bibitem{QZ10}
H. T. Quan, W. H. Zurek,  \href{https://doi.org/10.1088/1367-2630/12/9/093025}{New J. Phys. {\bf 12},  093025 (2010).}

\bibitem{scaleFermi}
Y. Nishida and D. Son, \href{https://doi.org/10.1103/PhysRevD.76.086004}{Phys. Rev. D {\bf 76}, 086004 (2007).}




\end{thebibliography}

\begin{thebibliography}{10}

\bibitem{CD96} Y. Castin and R. Dum, \href{https://doi.org/10.1103/PhysRevLett.77.5315}{Phys. Rev. Lett. {\bf 77}, 5315 (1996).}

\bibitem{KSS96} Yu. Kagan, E. L. Surkov, and G. V. Shlyapnikov, \href{https://doi.org/10.1103/PhysRevA.54.R1753}{Phys. Rev. A {\bf 54}, R1753(R) (1996).}

\bibitem{Gambardella75} P. J. Gambardella, \href{http://dx.doi.org/10.1063/1.522651}{J. Math. Phys.  {\bf 16}, 1172 (1975).}
\bibitem{Perelomov78} A. M. Perelomov, \href{http://dx.doi.org/10.1007/BF02156126}{Commun. Math. Phys. {\bf 63}, 9 (1978).}

\bibitem{GBD10} V. Gritsev, P. Barmettler, E. Demler,  \href{https://doi.org/10.1088/1367-2630/12/11/113005}{New J. Phys. {\bf 12}, 113005 (2010).}

\bibitem{delcampo11} A. del Campo
\href{http://dx.doi.org/10.1103/PhysRevA.84.031606}{Phys. Rev. A {\bf 84},  031606(R) (2011).}


\bibitem{Castin04} Y. Castin, \href{http://dx.doi.org/10.1016/j.crhy.2004.03.017}{Comptes Rendus Physique {\bf 5}, 407  (2004).}


\bibitem{StringariSTA}
D. Papoular and S. Stringari, \href{https://doi.org/10.1103/PhysRevLett.115.025302}{Phys. Rev. Lett. {\bf 115}, 025302 (2015).}

\bibitem{Lobo08}
C. Lobo and S. D. Gensemer, \href{https://doi.org/10.1103/PhysRevA.78.023618}{Phys. Rev. A {\bf 78}, 023618 (2008).}


\bibitem{LR69} H. R. Lewis and W. B. Riesenfeld, \href{http://dx.doi.org/10.1063/1.1664991}{J. Math. Phys. {\bf 10}, 1458 (1969).}


\bibitem{Menotti02} C. Menotti, P. Pedri, and S. Stringari, \href{https://doi.org/10.1103/PhysRevLett.89.250402}{Phys. Rev. Lett. {\bf 89}, 250402 (2002).}


\bibitem{Wu1}
S. Deng, P. Diao, Q. Yu, and H. Wu,  \href{https://doi.org/10.1088/0256-307X/32/5/053401}{Chin. Phys. Lett. {\bf 32}, 053401 (2015).}

\bibitem{Wu2}
S. Deng, Z. Shi, P. Diao, Q. Yu, H. Zhai, R. Qi, and H. Wu, \href{https://doi.org/10.1126/science.aaf0666 }{Science {\bf 353}, 371 (2016).}

\bibitem{Johnscale}
E. Elliott, J. Joseph, and J. Thomas, \href{https://doi.org/10.1103/PhysRevLett.89.250402}{Phys. Rev. Lett.\textbf{112}, 040405 (2014).}

\bibitem{Caoscience}
C. Cao, E. Elliott, J. Joseph, H.Wu, J. Petricka, T. Schaefer, and J. Thomas, \href{https://doi.org/10.1126/science.1195219}{Science  \textbf{331}, 58 (2011).}

\end{thebibliography}
\end{document}